\documentclass[
    ,final            
  ]
  {aipproc}

\layoutstyle{6x9}

\usepackage{bm}
\renewcommand{\vec}[1]{\boldsymbol{#1}}
\def\aap{\emph{Astron. Astrophys.}}
\def\apj{\emph{Astrophys. J.}}
\def\prl{\emph{Phys. Rev. Lett.}}
\def\pop{\emph{Phys. Plasmas}}
\def\jcp{\emph{J. Comput. Phys.}}
\def\jgr{\emph{J. Geophys. Res.}}
\def\mnras{\emph{Mon. Not. R. Astron. Soc.}}

\renewcommand{\div}{\nabla \cdot}
\newcommand{\rot}{\nabla \times}

\begin{document}

\title{Fluid and Magnetofluid Modeling of Relativistic Magnetic Reconnection}

\classification{52.27.Ny, 52.35.Vd, 95.30.Qd, 95.30.Sf}
\keywords      {Magnetic reconnection, relativistic plasmas}

\author{Seiji Zenitani}{
  address={NASA Goddard Space Flight Center}
}
\author{Michael Hesse}{
}
\author{Alex Klimas}{
}

\begin{abstract}
The fluid-scale evolution of relativistic magnetic reconnection is
investigated by using two-fluid and magnetofluid simulation models.
Relativistic two-fluid simulations demonstrate the meso-scale evolution beyond the kinetic scales,
and exhibit quasisteady Petschek-type reconnection. 
Resistive relativistic MHD simulations further show new shock structures
in and around the downstream magnetic island (plasmoid).  
General discussions on these models are presented.
\end{abstract}

\maketitle


\section{Introduction}

Magnetic reconnection is an important fundamental process in plasmas. 
It has drawn growing attention in extreme astrophysical sites such as
pulsars and magnetars,
where plasmas primarily consist of electron--positron pairs. 
One of the most notable applications is the magnetic dissipation problem in a pulsar wind,
an ultrarelativistic plasma flow from the pulsar magnetosphere. 
It is expected that an equatorial current sheet (like the heliospheric current sheet)
flaps extremely due to the oblique rotation of the central neutron star, and
that reconnection inside the current sheet dissipates the magnetic energy \cite{coro90}.
Importantly, in those environments the magnetic energy exceeds
the rest mass energy of lightweight electrons and positrons. 
When reconnection transfers magnetic energy to that of plasmas,
we need to take into account special relativity effects
both in the bulk motion and in the plasma heat.

Our understanding of relativistic reconnection is much more limited
than our understanding of the nonrelativistic counterpart.  
In the last decade, there has been some fundamental work
in the following two areas: 
magnetohydrodynamic (MHD) theories of a steady-state structure \cite{bf94b,lyut03,lyu05} and
time-dependent particle-in-cell simulations on kinetic scales
\cite{zeni01,claus04,zeni07,zeni08,lyu08}. 
Since these two are temporally and spatially separated,
it has been difficult to associate the results from the two research areas. 
After an initial attempt \cite{naoyuki06},
MHD simulations
of the basic reconnection process
have been stalled for a while.

In order to bridge the two areas, we have developed
two simulation models to study relativistic magnetic reconnection:
the electron-positron two-fluid model \cite{zeni09a,zeni09b} and
the resistive relativistic MHD (RRMHD) model \cite{zeni10b}.

\section{Two-fluid Simulations}

The basic equations consist of relativistic fluid equations for electrons and positrons,
and Maxwell equations \cite{zeni09a,zeni09b}.
For simplicity, $c$ is set to $1$.
\begin{eqnarray}
\label{eq:cont}
\partial_t ( \gamma_p n_p ) &=& -\div (n_p \vec{u}_p), \\
\label{eq:mom}
\partial_t \Big( { \gamma_p w_p \vec{u}_p } \Big)
&=& -\div \Big( { w_p \vec{u}_p\vec{u}_p } + \delta_{ij} p_p \Big)
+ \gamma_p n_p q_p (\vec{E}+\vec{v_p}\times\vec{B}) \nonumber \\
&&~~~~~~~~~~~~~~~~~~~~~~~~~~~~~~~~~~~~~~~~~~~~~~~~
- \tau_{fr} n_p n_e (\vec{u}_p-\vec{u}_e), \\
\label{eq:ene}
\partial_{t} \Big(\gamma_p^2 w_p - p_p \Big)
&=& -\div ( \gamma_p w_p \vec{u}_p ) + \gamma_p n_p q_p (\vec{v_p}\cdot\vec{E})
- \tau_{fr} n_p n_e ({\gamma}_p-{\gamma}_e),\\
\label{eq:cont2}
\partial_t ( \gamma_e n_e ) &=& -\div (n_e \vec{u}_e), \\
\label{eq:mom2}
\partial_t \Big( { \gamma_e w_e \vec{u}_e } \Big)
&=& -\div \Big( { w_e \vec{u}_e\vec{u}_e } + \delta_{ij} p_e \Big)
+ \gamma_e n_e q_e (\vec{E}+\vec{v_e}\times\vec{B}) \nonumber \\
&&~~~~~~~~~~~~~~~~~~~~~~~~~~~~~~~~~~~~~~~~~~~~~~~~
- \tau_{fr} n_p n_e (\vec{u}_e-\vec{u}_p), \\
\label{eq:ene2}
\partial_{t} \Big(\gamma_e^2 w_e - p_e \Big)
&=& -\div ( \gamma_e w_e \vec{u}_e ) + \gamma_e n_e q_e (\vec{v_e}\cdot\vec{E})
- \tau_{fr} n_p n_e ({\gamma}_e-{\gamma}_p), \\
\partial_t {\vec{B}} &=& - \rot \vec{E},
~~~~~~~~
\partial_t {\vec{E}} = \rot \vec{B} - 4\pi ( q_p n_p \vec{u}_p+q_e n_e \vec{u}_e ).
\end{eqnarray}
In these equations,
the subscript $p$ means positron properties (and $e$ for electrons),
$\gamma$ is the Lorentz factor,
$n$ is the proper density,
$\vec{u}=\gamma\vec{v}$ is the 4-vector,
$w$ is the enthalpy,
$p$ is the proper pressure, and $q_p=-q_e$ is the charge.
We assume a $\Gamma$-law equation of state
with the adiabatic index $\Gamma=4/3$.
Therefore the enthalpy is given by $w=nmc^2+[\Gamma/(\Gamma-1)]p$.
In order to mimic the effective resistivity,
we introduce an inter-species friction force
in the last terms of the momentum and energy equations with a coefficient $\tau_{fr}$.
Internally, we consider the energy density
relative to the rest mass energy density
(i.e., we subtract Eq.~\eqref{eq:cont} from Eq.~\eqref{eq:ene})
similarly as in earlier work \cite{naoyuki06,marti03}.

The time evolution is solved by a standard numerical scheme,
i.e. a modified Lax--Wendroff scheme. 
A common difficulty of relativistic fluid simulations is that
the fluid macro properties (the ``conservative variables'') on the left hand sides
are nonlinear combinations of the basic elements (the ``primitive variables'')
such as $\gamma$ and $p$ \cite{marti03}. 
In our simulation, 
we calculate the primitive variables from the conservative variables
by analytically solving a quartic equation of $|{u}|$ \cite{zeni09a}
on all grid cells at each half timestep.

We investigate a two-dimensional system evolution in the $x$--$z$ plane.
The reconnection point is set to the origin $(x,z)=(0,0)$.
We employ a Harris-like initial configuration:
$\vec{B} = B_0 \tanh(z)~\vec{\hat{x}},$
$\vec{j} = {B_0} \cosh^{-2}(z)~\vec{\hat{y}}$,
$n = n_0 \cosh^{-2}(z) + 0.1 n_{0}$, $p=nmc^2$,
$\vec{u}=0$, and $\vec{E} = \eta \vec{j}$. 
In the upstream region ($|z|{\gg}0$),
the magnetization parameter is $\sigma = B_0^2/[4\pi (2w)] =4$.
The relevant Alfv\'{e}n speed is $c_{A,up}=[\sigma/(1+\sigma)]^{1/2} \sim 0.89c$
or $\gamma c_{A,up} = \sqrt{\sigma} = 2$.
We assume that the effective resistivity $\eta \propto \tau_{fr}$ is
localized around the reconnection point.
Neumann-like boundaries are located at $x = \pm 120$ and at $z = \pm 60$.

\begin{figure}
  \includegraphics[width=\columnwidth]{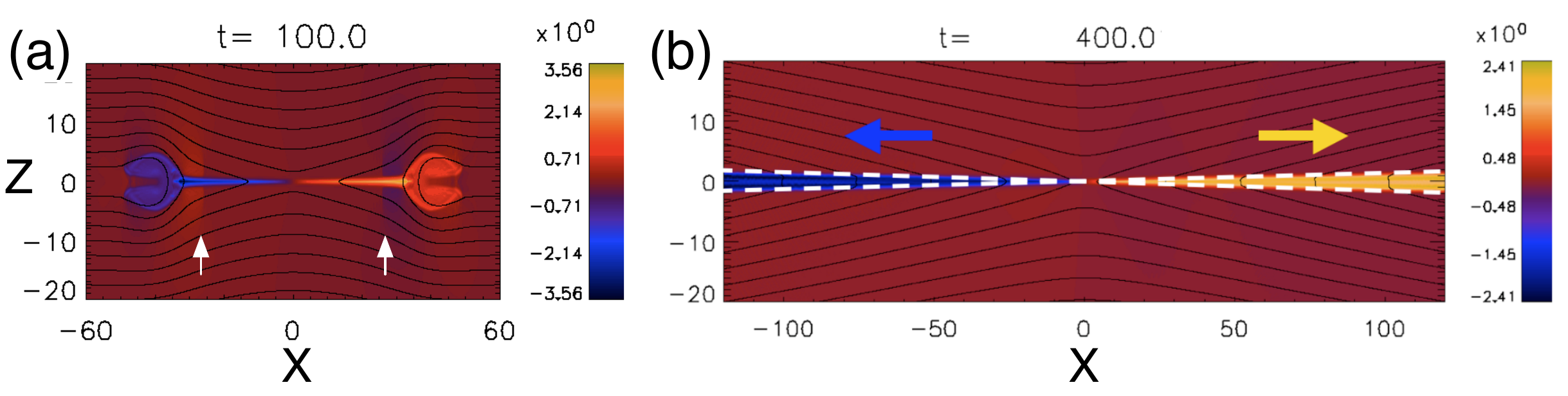}
  \caption{(Color online)
The plasma 4-velocity $u_x$ and the magnetic field lines
at (a) $t=100$ and (b) $t=400$.
In Panel (a) the arrows indicate the vertical structures,
which will be discussed later.
In Panel (b) the dashed lines represent Petschek-type slow-shock regions.
(From run L3 in Ref.~\cite{zeni09a})
}
\end{figure}

Reconnection takes place around the center.
Figure 1a shows the $x$-component of the 4-velocity, $u_x=\gamma v_x$, at $t=100$
in unit of the light transit time $c^{-1}$.
Magnetic field lines are transported from the background regions,
cut and reconnected at the reconnection point, and then
ejected with the reconnection jets. 
One can see the bi-directional jets from the reconnection point and
magnetic islands (plasmoids) in front of the jets. 
The 4-velocity of the reconnection jet is typically $u_x \sim 2$,
which is Alfv\'{e}nic with respect to the upstream condition, 
and it is $u_x \sim 3.5$ at the local maximum right behind the plasmoid. 
Reconnection sets up an inflow speed of $v_z \sim 0.14$-$17$.
Since the reconnection rate or a normalized form of the flux transfer speed
is an order of $\mathcal{R} \sim (v_{in}/v_{out}) \sim O(0.1)$,
this is a fast reconnection. 
Later, the plasmoids reach and go through
the open boundaries at $x = \pm 120$.
Although there are minor reflections from the boundaries, 
the system exhibits a quasisteady structure in the long term.
Figure 1b presents the late-time snapshot at $t=400$.
The reconnection jets are confined by slow-shock-like regions,
as indicated by the dashed lines in Fig. 1b.
One can see a quasisteady Petschek-type structure \cite{petschek}. 

Interestingly, our analysis reveals that
the Petschek outflow becomes narrower than the nonrelativistic counterpart,
as theoretically predicted by Lyubarsky \cite{lyu05}.
Meanwhile, reconnection proceeds faster,
especially in the ultrarelativistic, high-$\sigma$, regimes.
It was also found that an out-of-plane magnetic field (guide field)
drastically changes the composition of the energy outflow.
Without a guide field the main energy carrier is the plasma enthalpy flux, while
the Poynting flux carries the energy in the presence of a moderate guide-field \citep{zeni09b}.

\section{Resistive MHD Simulations}

We use the following RRMHD equations \cite{naoyuki06,kom07}
in Lorentz--Heaviside units with $c=1$.
\begin{eqnarray}
\partial_t (\gamma \rho) + \div (\rho \vec{u}) &=& 0, \\
\partial_t ( \gamma w\vec{u} + \vec{E}\times\vec{B} )
+ \div \Big( ( p + \frac{B^2+E^2}{2} ) \vec{I}
+ w \vec{u}\vec{u}
- \vec{B}\vec{B} - \vec{E}\vec{E} \Big)
&=& 0, \\
\partial_t ( \gamma^2w-p + \frac{ {B}^2+{E}^2 }{2} )
+ \div ( \gamma w\vec{u} + \vec{E}\times\vec{B} ) &=& 0, \\
\partial_t\vec{B} + \nabla \times \vec{E} = 0, ~~~~~~
\partial_t\vec{E} - \nabla \times \vec{B} = -\vec{j}, ~~~~~~
\partial_t{\rho_c} + \div \vec{j} &=& 0,
\\
\label{eq:ohm}
\gamma \Big( \vec{E} + \vec{v}\times\vec{B} - (\vec{E}\cdot\vec{v}) \vec{v} \Big)
= \eta ( \vec{j} - \rho_c \vec{v} )&&
\end{eqnarray}
Here, $\rho_c$ is the charge density.
We developed three RRMHD codes \cite{naoyuki06,kom07,pal09}.
Among them, the present simulations are carried out by the HLL-type code \cite{kom07}.
The RRMHD schemes differ from nonrelativistic schemes in the following two ways.
One is related to the relativistic primitive variables.
We use the same quartic equation solver to recover the primitive variables.
The other involves the usage of Amp\`{e}re's law. 
In the nonrelativistic MHD, we assume $\vec{j}\equiv\rot \vec{B}$ and then
derive $\vec{E}$ from Ohm's law.
In the relativistic MHD, we use the Amp\`{e}re's law to advance $\vec{E}$.
Configurations are similar to those in the two-fluid case.
We employ left-right symmetry to reduce computational cost. 
We use a localized resistivity $\eta=\eta(x,z)$
in the relativistic Ohm's law (Eq. \ref{eq:ohm}).

The system evolves similarly in the two-fluid and MHD runs.
In the present case, we can resolve sharper shock structures
because we employ a shock-capturing scheme.
In the two-fluid model,
the system contains the Larmor radius or the inertial length
and so
it makes no sense to discuss shorter structures. 
We have no such concern in the scale-free RRMHD model.
Figure 2a
shows the $u_x$-profile in the developed stage. 
This is a Petschek-type reconnection geometry.
The typical outflow is Alfv\'{e}nic, $u_x \sim 2$.
The jet hits the plasmoid at a fast shock (``FS'' in Fig. 2a)
at the maximum 4-velocity of $u_x \sim 2.75$.

\begin{figure}
  \includegraphics[width=\columnwidth]{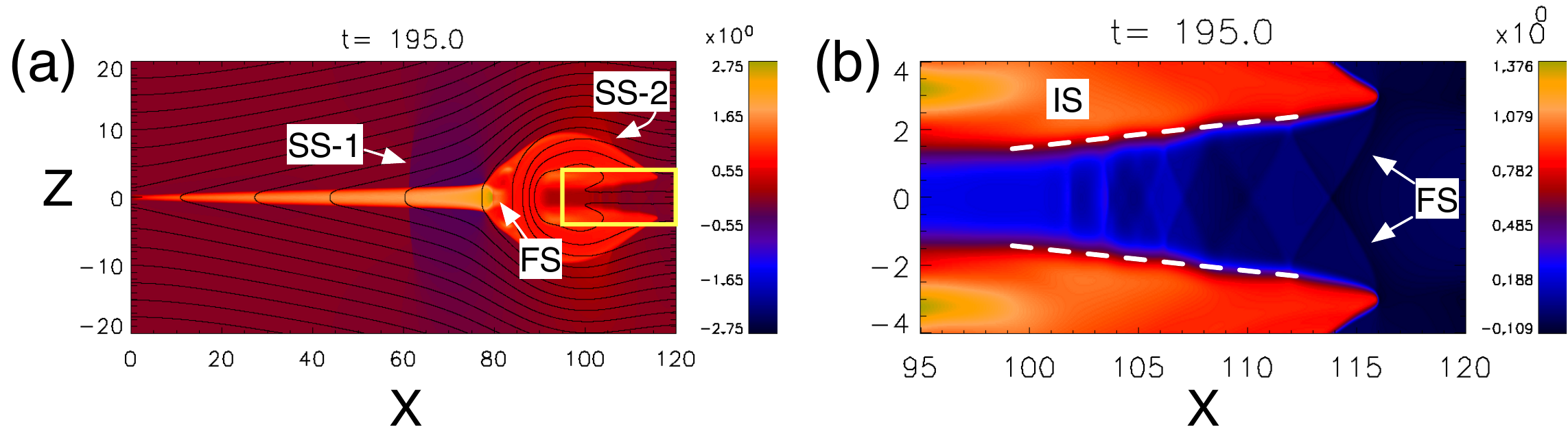}
  \caption{(Color online)
(a) The $x$-component of 4-velocity ($u_x$) at the developed stage.
The contour lines indicate the magnetic field lines.  
(b) The $u_x$-profile in the front side of the plasmoid.
The domain is identical to the yellow box in Panel (a).
(From run 1 in Ref.~\cite{zeni10b})
}
\end{figure}

The simulation reveals new shock structures:
postplasmoid vertical slow shocks,
forward vertical slow shocks,
and the shock-reflection structure. 
One can see the postplasmoid shocks at $x\sim 60$,
as indicated by ``SS-1'' in Fig. 2a. 
The shock propagates in the $+x$-direction
at the front of the reverse plasma flow
($u_x<0$; blue regions in Fig. 2a).
As the plasmoid moves,
it compresses the surrounding plasmas on both upper and lower sides, 
and the cavity region also appears behind the plasmoid.
The pressure gradient between those two regions drives
the reverse flow along the field lines
\cite{zeni10b,zeni11}.
The right side is the shock upstream.
One can also recognize similar shock-like structure
in the two-fluid run, as indicated by arrows in Fig. 1a.
We find another small shock outside the plasmoid
(``SS-2'' in Fig. 2a; see also Fig. 2c in Ref.~\cite{zeni10b}).
This is a relativistic version of
the forward vertical slow-shock \cite{zeni11}.
We further analyze the properties of these two shocks.
In the shock normal direction \cite{son67},
we evaluate the shock speed $v_{shock}$ and the MHD wave speeds
in the lab-frame \cite{keppens08}.
Key results are presented in Table \ref{table}.
In both cases, on the right (upstream) sides,
the shocks travel faster than the outgoing slow-mode, but slower than the Alfv\'{e}n mode,
$v_{s+,R} < v_{shock} < c_{A+,R}$.
This is the character of slow shocks. 
Finally,
we find the shock-reflection structure in the front side,
as indicated by the small box in Fig. 2a.
Figure 2b shows the $u_x$-profile in the same domain.
This one takes advantage of the current sheet configuration.
Since there is initially a dense plasma near the center $z \sim 0$,
plasma flows are bifurcated.
The twin flows and the central immobile plasmas are separated by
intermediate shocks \cite{shuei01} (the dashed lines or ``IS''). 
As seen in Fig. 2b, the twin edges invoke oblique bow shocks (fast shocks: ``FS'' in Fig. 2b),
and then they are reflected by the intermediate shocks many times.
In the long term, the system exhibits a chain of diamond-shaped structures. 
Inside the central immobile region, the magnetic fields are weak or zero.
The fastest wave is the sound speed in the current sheet $c_{s,cs} \sim 0.53$.
The twin edges move faster $\sim 0.75 > c_{s,cs}$, because
they are essentially driven by the reconnection jet $\sim c_{A,up} \sim 0.89$.
We therefore expect the oblique shocks when $c_{A,up} > c_{s,cs}$.

\begin{table}
\begin{tabular}{lccccccc}
\hline
&
Location &
$(n_x,n_z)$ &
$v_{shock}$ &
$v_{s+,L}$ &
$v_{s+,R}$ &
$c_{A+,R}$ &
\\
\hline
SS-1 &
(61.5, 5.0) &
(1.00,-0.08) &
0.41 &
0.50 &
0.29 &
0.80
\\
SS-2 &
(105.9, 8.2) &
(0.90,0.43) &
0.62 &
0.66 &
0.56 &
0.80
\\
\hline
\end{tabular}
\caption{Selected shock properties.}
\footnotetext{
The shock normal vector $\vec{\hat{n}}$,
the shock velocity $v_{shock}$,
the slow-mode, and the Alfv\'{e}n wave speeds
on the left ($L$) and right ($R$) sides of the shock.
All velocities are evaluated 
in the lab-frame
in the $\vec{\hat{n}}$-direction.
}
\label{table}
\end{table}

\section{Discussion}

These simulations have revealed new insights in
the fluid-scale properties of relativistic reconnection.
The relativistic Petschek reconnection is fast,
it features an Alfv\'{e}nic outflow jet, and
the outflow exhaust becomes narrower and narrower
as the magnetization parameter $\sigma$ increases.
The overall results confirm Lyubarsky's theory \cite{lyu05}. 
In our simulations, $\sigma$ ranges up to ${\sim}10$.
Extended parameter surveys and many basic issues need to be investigated. 

The simulations exhibit new shock structures,
especially in the RRMHD model.
We think that they appear when the Alfv\'{e}nic reconnection jet exceeds
the sound speeds \cite{zeni10b,zeni11}. 
Comparing the outflow speed $\approx c_{A,up}=[\sigma/(1+\sigma)]^{1/2}$ and
the upper limit of the sound speed $1/\sqrt{3}$,
a sufficient condition for a supersonic reconnection jet is
\begin{equation}
\sigma > 1/2.
\end{equation}
Therefore we expect various MHD shocks in the reconnection system
in high-$\sigma$ regimes.

At present, the most important issue is the effective resistivity,
which crucially controls the system evolution.
In this work, we have employed a spatially-localized resistivity for our main results.
It is known that such resistivity leads to a Petschek-type fast reconnection
in the nonrelativistic MHD studies \cite{ugai77}. 
Meanwhile, other resistivity models give significantly different pictures:
a turbulent outflow with secondary islands \cite{zeni09a},
a slow Sweet--Parker reconnection and so on \cite{zeni10b}.
We need to find a practical resistivity model
to reproduce realistic reconnection,
by referring to the kinetic results \cite{zeni07}.
Of course a true resistivity is desirable, but
this is a long-standing problem in reconnection physics.

The RRMHD model has another problem to overcome.
The standard equations are very stiff
when the magnetic Reynolds number $S \sim 1/\eta$ is high.
Implicit-type schemes \cite{pal09,dumbser09}
have been recently developed to explore high-$S$ regimes
and
applications to Sweet--Parker reconnection are in progress \cite{takahashi10}.

The development of the two-fluid and RRMHD models will benefit
the large-scale modeling of relativistic MHD instabilities \cite{ober09,ober10}.
Even in ideal MHD studies,
the late-time evolution often exhibits
potential reconnection sites with antiparallel magnetic fields
or (numerical) reconnection.
Reconnection with a physical resistivity will be a basic element of
these important problems.

Our results have implications for space physics also.
Inspired by these results,
we carried out nonrelativistic MHD simulations and
found the same shock structures in low-beta plasmas \cite{zeni11}.
In other words, our discoveries are ubiquitous
in both relativistic and nonrelativistic reconnections. 
In the solar corona, 
the fast-shock-reflection structure at the plasmoid front
may be related to energetic particle acceleration.

\begin{theacknowledgments}
The authors acknowledge valuable discussions with
T. Miyoshi, Y. Mizuno, H.~R. Takahashi, and A.~F. Vinas.
This research was supported by the NASA Center for Computational Sciences.
One of the authors (S.Z.) gratefully acknowledges support from
NASA's postdoctoral program and
JSPS Fellowship for Research Abroad.
\end{theacknowledgments}

\bibliographystyle{aipproc}   

\begin{thebibliography}{20}

\bibitem{coro90}
F. V. Coroniti,
\apj, {\bf 349}, 538--545 (1990).

\bibitem{bf94b}
E.~G. Blackman, \& G. B. Field, \prl, {\bf 72}, 494--497 (1994).

\bibitem{lyut03}
M. Lyutikov, \& D. {Uzdensky}, \apj, {\bf 589}, 893--901 (2003).

\bibitem{lyu05}
Y. Lyubarsky, \mnras, {\bf 358}, 113--119 (2005).

\bibitem{zeni01}
S. Zenitani, \& M. Hoshino,
\apj, {\bf 562}, L63--L66 (2001).

\bibitem{claus04}
C. H. Jaroschek, R. A. Treumann, H. Lesch, \& M. Scholer,
\pop, {\bf 11}, 1151--1163 (2004).

\bibitem{zeni07}
S. Zenitani, \& M. Hoshino,
\apj, {\bf 670}, 702--726 (2007).

\bibitem{zeni08}
S. Zenitani, \& M. Hoshino,
\apj, {\bf 677}, 530--544 (2008).

\bibitem{lyu08}
Y. Lyubarsky, \& M. Liverts,
\apj, {\bf 682}, 1436--1442 (2008).

\bibitem{naoyuki06}
N. Watanabe, \& T. Yokoyama,
\apj, {\bf 647}, L123--L126 (2006).

\bibitem{zeni09a}
S. Zenitani, M. Hesse, \& A. Klimas,
\apj, {\bf 696}, 1385--1401 (2009).

\bibitem{zeni09b}
S. Zenitani, M. Hesse, \& A. Klimas,
\apj, {\bf 705}, 907--913 (2009).

\bibitem{zeni10b}
S. Zenitani, M. Hesse, \& A. Klimas,
\apj, {\bf 716}, L214--L218 (2010).

\bibitem{marti03}
J.~M. Mart{\'{\i}}, \& E. M{\"u}ller,
{\itshape Living Reviews in Relativity}, {\bf 6}, 7 (2003).

\bibitem{petschek}
H.~E. Petschek, in AAS/NASA Symposium on the Physics of Solar Flares, Magnetic Field Annihilation, ed. W. N. Ness (NASA: Washington, DC), pp. 425--439 (1964).

\bibitem{kom07}
S.~S. Komissarov,
\mnras, {\bf 382}, 995--1004 (2007).

\bibitem{pal09}
C. Palenzuela, L. Lehner, O., Reula, \& L. Rezzolla,
\mnras, {\bf 394}, 1727--1740 (2009).

\bibitem{zeni11}
S. Zenitani, \& T. Miyoshi,
\pop, {\bf 18}, 022105 (2011).

\bibitem{son67}
B. U. O. Sonnerup, \& L. J. Cahill Jr.,
\jgr, {\bf 72}, 171--183 (1967).

\bibitem{keppens08}
R. Keppens, \& Z. Meliani,
\pop, {\bf 15}, 102103 (2008).

\bibitem{shuei01}
S.~A. Abe, \& M. Hoshino,
{\itshape Earth Planets Space}, {\bf 53}, 663--671 (2001).

\bibitem{ugai77}
M. Ugai, \& T. Tsuda, {\itshape Journal of Plasma Physics}, {\bf 17}, 337--356 (1977).

\bibitem{dumbser09}
M. Dumbser, \& O. Zanotti, \jcp, {\bf 228}, 6991--7006 (2009).

\bibitem{takahashi10}
H.~R. Takahashi, J. Matsumoto, Y. Masada, \& T. Kudoh,
in {\it Deciphering the Ancient Universe with Gamma-ray Bursts},
ed. N. Kawai and S. Nagataki, AIP Conf. Proc. 1279,
American Institute of Physics, Mellville, New York, pp. 427--429 (2010).

\bibitem{ober09}
M. Obergaulinger, P. {Cerd{\'a}-Dur{\'a}n}, E. {M{\"u}ller}, \& M.~A. Aloy,
\aap, {\bf 498}, 241--271 (2009).

\bibitem{ober10}
M. Obergaulinger, \& M.~A. Aloy,
\aap, {\bf 515}, A30 (2010).

\end{thebibliography}

\end{document}